\documentclass[vecphys]{svmult}

\usepackage{makeidx}         % allows index generation
\usepackage{graphicx}        % standard LaTeX graphics tool
\usepackage{amsmath}
\usepackage{amssymb}
\usepackage{multicol}        % used for the two-column index
\usepackage[bottom]{footmisc}% places footnotes at page bottom
\newcommand{\kc}{\mathcal{K}_C}
\newcommand{\krho}{\mathcal{K}_\rho}
\newcommand{\kp}{\mathcal{K}_P}
\newcommand{\kb}{\mathcal{K}_B}
\newcommand{\evec}{\mathbf{e}}
\newcommand{\kpe}{\vec{k}_\perp}
\newcommand{\kpa}{\vec{k}_\parallel}
\newcommand{\ep}{\vec{e}_\parallel}
\newcommand{\el}{\vec{e}_l}
\newcommand{\ea}{\vec{e}_A}
\newcommand{\er}{\vec{e}_r}
\newcommand{\etal}{\textit{et al.\ }}
\begin{document}

\title*{Pressure-driven instabilities in astrophysical jets}
\author{Pierre-Yves Longaretti}
\institute{Laboratoire d'Astrophysique de Grenoble (LAOG)\\
CNRS and Universit\'e Joseph Fourier\\
BP 53 38041 Grenoble Cedex 9 -- France\\
\texttt{Pierre-Yves.Longaretti@obs.ujf-grenoble.fr}}
\maketitle

The enormous distances over which astrophysical jets propagate
without losing their coherence certainly constitute one of the
most striking features of these objects. Typically, jets from
Young Stellar Objects (hereafter YSOs) do reach out to a few
parsecs, while the radial extent of their region of origin appears
to be smaller than $\sim 100$ A.U, making jets extremely elongated
structures.

Blandford and Rees \cite{BR74} already pointed out in their 1974
pioneering work that in laboratory experiments, jets do not
propagate much farther than about ten times their radii, which
makes the propagation lengths of astrophysical jets all the more
impressive. The simplest way out of this conundrum would be to
assume that jets are ballistic. Indeed, for YSO jets at least, the
observed opening angle ($\sim 5^\circ$) is consistent with the
idea that they freely expand when one compares their thermal and
bulk velocities. However, this option leaves open the issue of the
formation of such powerful jets in the first place. And, as
critically, the ballistic hypothesis does not explain how these
jets survive the development of the Kelvin-Helmholtz instability,
which is now known to be quite disruptive in purely hydrodynamic
jets \cite{Bod95} \cite{Bod98}.

The shortcomings of the simple ballistic picture certainly
motivated to some extent the elaboration of MHD jet models. Such
models, however, are also prone to instabilities. The most
important ones discussed in the literature can be grouped into
three categories:
\begin{itemize}
  \item MHD Kelvin-Helmholtz instability. As for its HD
  counterpart, the driving agent is the velocity gradient at the
  jet/external medium interface. This instability has received a
  lot of attention in the literature, as the largest source of
  free energy in a jet is its bulk motion (see \cite{Bir91} and
  \cite{F98} for reviews).
  \item Conversely, the presence of a magnetic field provides a
  source of instability even in the absence of bulk motion.
  Ideal MHD instabilities are commonly divided into current- and
  pressure-driven, according to the driving factor (equilibrium
  parallel current in the first case, and gas pressure versus
  field-line curvature in the second).
  \item Radiative instabilities, related to the coupling of the
  radiation field and of the plasma dynamical quantities.
\end{itemize}

The structure of jets is not precisely known, which is one of the
difficulties in analyzing their stability. Most stability analyzes
assume that jets can be described as some sort of cylindrical
column in motion, pervaded by a magnetic field. Self-collimated
jet models are not exactly cylindrical, but as the observed
opening angles are small, the assumption of cylindrical shape is
not expected to be a major limitation.

More critically, such jet models have a helical field structure,
with the azimuthal component of the magnetic field dominating over
the vertical one in the outer jet regions. This follows in most
models because the magnetic tension is the confining force
ensuring self-collimation. Static MHD columns (i.e., not subject
to the bulk motion characterizing MHD jets) pervaded by a helical
magnetic field are referred to as ``screw pinches" in the fusion
literature. It is also known in this context that the dominance of
the azimuthal field component leads to both types of MHD
instabilities mentioned above, and may cause the disruption of the
plasma column itself on a few dynamical time-scales. This has long
been an argument against magnetically self-confined jet models.
However, recent investigations indicate that a bulk motion can
play an important stabilizing role (see section 5 for
pressure-driven instabilities, and e.g. \cite{ALB00} and
references therein for current-driven instabilities). Conversely,
the presence of a magnetic field can help stabilizing the
Kelvin-Helmholtz modes \cite{RJF00}, \cite{BK02}. These recent
advances seem to indicate that a sophisticated equilibrium jet
structure is required if one is to understand jet stability
properties, a state of development not yet reached by the subject,
but that now appears to be within sight.

To conclude these introductory remarks, I would like to point out
that, in the nonlinear phase, an instability can have three broad
types of outcome: i/ disruption of the fluid configuration (in the
case at hand, of the jet as a jet); ii/ internal reorganization,
the flow becoming laminar again in the end; iii/ turbulence (with
or without internal reorganization of the structure). The most
prominent objective of the study of jet stability is to understand
how the first issue is avoided in real jets; this issue may well
be seen as our inability to formulate the initial value problem
correctly. A second but important issue is to understand how
turbulence might be driven by jet destabilization. This issue is
probably more important in AGN than in YSO jets, as turbulence is
often invoked in the former context as a source of high energy
particle acceleration.

The object of these lecture notes is pressure-driven
instabilities. As most investigations of this problem have been
made in the fusion context for static columns, this essential
aspect of the subject will first be reviewed, before briefly
presenting the more recent (and more scant) results on moving
columns. The next section presents some general ideas about the
physical origin of MHD instabilities; the concept of magnetic
shear is introduced there as well, and its stabilizing role on
pressure-driven instabilities, expressed by Suydam criterion, is
discussed. Section \ref{idealMHD} introduces the Lagrangian form
of the perturbation equations used in static columns, as this
formulation is the most powerful to derive general results, such
as those derived from the Energy Principle presented in section
\ref{EnPrin}. Section \ref{disprel} presents the dispersion
relation of pressure-driven instabilities in low magnetic shear
that are expected to characterize jets outer regions. The few
published results on moving columns (i.e., jets) are presented in
section \ref{movcol}. The last section summarizes the present
state of understanding of this aspect of jet stability, and
outlines areas where improvement is needed.

Sections \ref{sec:gen} and \ref{movcol} are intended for a general
audience, while sections \ref{idealMHD}, \ref{EnPrin} and
\ref{disprel} are more theoretical in nature. The exposition is
aimed at the graduate student level.

\section{Heuristic description of MHD instabilities}
\label{sec:gen}

This first section is intended to provide the reader with some
qualitative and semi-quantitative ideas about the onset and
characteristics of pressure-driven instabilities, leaving
technical aspects of the stability analysis to the later sections.

\subsection{Qualitative conditions of ideal MHD instabilities in static equilibria}
\label{sec:base}

The equilibrium configurations leading to an ideal MHD instability
have been well investigated in the fusion literature. For
current-driven instabilities, the first criterion was devised by
Kruskal and Shafranov. Basically, it states that in cylindrical
column of length $L$, instability follows if the magnetic field
line rotates more than a certain number of times around the
cylinder, from end to end. The exact number of rotations required
for instability is dependent on the considered equilibrium
configuration; it is usually of order unity.

Concerning pressure-driven instabilities, a more clear-cut
necessary condition of instability can be stated: instability
follows once the pressure force pushes the plasma outwards from
the inside of the field line curvature. This condition can be
derived from the Energy Principle, as will be shown later on.

\begin{figure}
\centering
\includegraphics[scale=0.5]{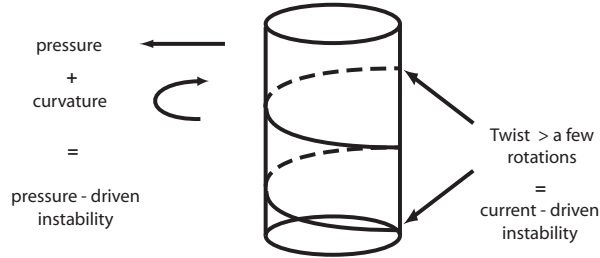}
\caption{Qualitative description of the conditions of onset of MHD
instabilities (see text for details).} \label{MHD-Inst}
\end{figure}

These conditions of onset of instability are illustrated on
Fig.~\ref{MHD-Inst}. In an actual plasma, the origin of an
instability (current- or pressure-driven) is usually not easy to
pinpoint except in special instances. For example, if the plasma
is cold (no pressure force), the instability is necessarily
current-driven. Also the growth rates of current-driven modes are
known to decrease with spatial order -- e.g., they decrease with
increasing azimuthal wave-number $m$ -- while the most unstable
pressure-driven modes have a growth rate which is nearly
independent on the wavenumber. Consequently, large wavenumber
unstable modes are therefore always pressure-driven in a static,
ideal MHD column. Besides these two limiting cases, an MHD
instability almost always results from an inseparable mix of
pressure and current driving. If the column is moving, the
distinction between Kelvin-Helmholtz, current- and pressure-modes
is even more blurred, except in some cases, where branches of
instability can be identified by taking appropriate limits.

In terms of outcome of the instability, it is essential to know
whether unstable modes are internal or external, i.e., have
vanishing or substantial displacement on the plasma surface (here,
the jet surface) . It is well-known in the fusion context that
unstable external modes are prone to disrupt the plasma, as may be
the case, e.g., with the $m=1$ (``kink") current-driven mode.

In the next sections, mostly high wavenumber modes will be
examined, where the pressure-driving is most obvious, in order to
best identify the characteristic features of this type of
instability.

\subsection{Magnetic shear, magnetic resonances, and Suydam's
criterion}\label{suyd}

\begin{figure}
\centering
\includegraphics[scale=0.5]{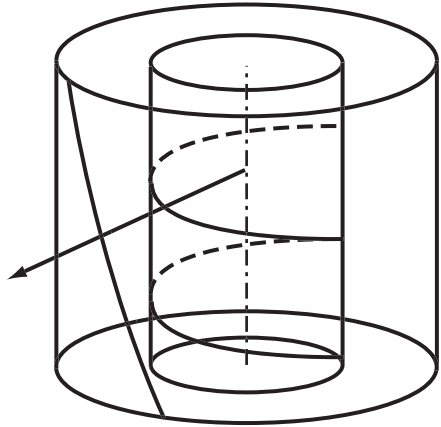}
\caption{The change in the pitch of field lines between magnetic
surfaces is the source of the magnetic shear (see text).}
\label{shearfig}
\end{figure}

The concept of magnetic shear plays an important role in the
understanding of the stability of pressure-driven mode. The
magnetic shear characterizes the change of orientation of field
lines when moving perpendicularly to magnetic surfaces. In the
case of cylindrical equilibria, this concept is illustrated on
Fig.~\ref{shearfig}. Magnetic surfaces are cylindrical. Field
lines within magnetic surfaces have an helix shape; the change of
helix pitch $rB_z/B_\theta$, characterizes the magnetic shear. A
quantity related to the pitch and largely used in the fusion
community is the safety factor $q$:

\begin{equation}\label{safety}
  q=\frac{r B_z}{R_o B_\theta},
\end{equation}

\noindent where $R_o$ is the column radius. For reasons soon to be
discussed, a high enough safety factor is required for stability,
hence its name. The magnetic shear $s$ is defined as

\begin{equation}\label{shear}
  s\equiv \frac{r}{q}\frac{dq}{dr}.
\end{equation}

Magnetic resonances constitute another important key to the
question of stability. A cylindrically symmetric equilibrium is
invariant in the vertical and azimuthal direction, so that
perturbations from equilibrium can without loss of generality be
expanded in Fourier terms in these directions and assumed to be
proportional to $\exp i (m\theta+kz)$. Magnetic resonances are the
(cylindrical) surfaces where the wave vector ${\vec k} = m/r {\vec
e}_\theta + k {\vec e}_z$ is perpendicular to the equilibrium
magnetic field:

\begin{equation}\label{magres}
  \frac{{\vec k}\cdot{\vec B}_o}{B_o}\equiv k_\parallel=
\frac{1}{B_o}\left(\frac{m}{r}B_\theta+k B_z\right)=0
\end{equation}

\noindent where $k_\parallel$ is the component of the wave vector
parallel to the equilibrium magnetic field. The significance of
these surfaces stems from the fact that in general, dispersion
relations incorporate a stabilizing piece of the form $V_A^2
k_\parallel^2$, where $V_A$ is the Alfv\'en speed. This term is
responsible for the propagation of Alfv\'en waves, and arises from
the restoring force due to the magnetic tension (see section
\ref{disprel} for the precise meaning of these statements). As
such, it is always stabilizing. Obviously, this stabilization is
minimal in the vicinity of a magnetic resonance for a given
($m,k$) mode, so that pressure-driven instabilities are
preferentially triggered at magnetic resonances for any given
mode.

Note however that a large magnetic shear limits the role of
magnetic resonances in the destabilization of the plasma. Indeed,
defining the perpendicular wave number

\begin{equation}\label{kperp}
  k_\perp=-\frac{1}{B_o}\left(\frac{m}{r}B_z - k B_\theta\right),
\end{equation}

\noindent and designating by $r_c$ the radial position of the
magnetic resonance of the ($m,k$) mode, one finds that

\begin{equation}\label{kpar}
  k_\parallel\simeq \frac{B_\theta B_z}{B_o^2} k_\perp s \frac
{r-r_c}{r_c},
\end{equation}

\noindent to first order in $(r-r_c)/r_c$ in the vicinity of the
magnetic resonance $r_c$. This implies that $V_A k_\parallel$ will
remain small either if $s \ll 1$ (small shear) or if the magnetic
field is mostly perpendicular ($|B_\theta| \ll |B_z|$) or
azimuthal ($|B_z| \ll |B_\theta|$) so that $|B_\theta B_z|/B_o^2
\ll 1$. However, if the field is predominantly vertical, it is
little curved, and pressure destabilization is expected to be weak
or non-existent according to the description of the condition on
instability depicted in Fig.~\ref{MHD-inst}; furthermore, $s$
being a logarithmic derivative is usually of order unity.
Therefore, in practice stabilization by magnetic tension will be
reduced essentially when the field is mostly azimuthal.

These features are embodied in Suydam criterion, which expresses a
sufficient condition for instability:

\begin{equation}\label{suydam}
  \frac{B_z^2}{8\mu_o r}s^2+\frac{dP}{dr} > 0.
\end{equation}

\noindent The converse of this statement is a necessary condition
for stability. The origin of this criterion is briefly discussed
in section \ref{EnPrin}. It turns out that this condition is both
a necessary and sufficient condition of instability for large
wavenumber modes \cite{DTYNM04}. The condition of instability
requires $dP/dr < 0$, which agrees with our heuristic description
of the onset of instability given above. It will be also discussed
in Section \ref{EnPrin} that the growth rates $\gamma$ of
pressure-driven instabilities are $\gamma\sim C_S/R_o$ ($C_S$ is
the sound speed and $R_o$ the jet radius).

Coming back to Eq.~(\ref{suydam}), the first term is stabilizing,
but the stabilization will be minimal in the condition just
discussed, i.e., when the field is mostly azimuthal. Indeed, in
this case, the equilibrium condition Eq.~(\ref{radequi}) implies
that $dP/dr\sim B_\theta^2/\mu_o r \gg B_z^2/r\sim B_z^2 s^2/r$.
This situation is expected to hold in magnetically self-confined
jets outer regions. Indeed, most such jet models (e.g., \cite{L96}
and \cite{F97}) have $|B_\theta| \gg |B_z|$ in the asymptotic jet
regime to ensure confinement. This feature combined to the
previous statement that MHD instabilities involving the boundary
are most prone to disrupt static MHD columns makes the assessment
of the role of pressure-driven instabilities in MHD jets
particularly critical for the viability of such models. This
viability hinges on the hopefully stabilizing role of the jet bulk
motion (see section \ref{movcol}).

\section{Ideal MHD in static columns:}\label{idealMHD}

The simplest framework in which the stability of jets can be
investigated is ideal magnetohydrodynamics (MHD). Justifications
and limitations of this approach are briefly discussed in Appendix
\ref{app:idealmhd}.

\subsection{Equations}

The MHD equations used in these notes are the continuity equation,
the momentum equation without the viscous term, the induction
equation without the resistive term, and a polytropic equation of
state. Incompressibility is not assumed, as pressure-driven modes
are not incompressible except at the marginal stability limit.
These equations read

\begin{equation}\label{cont}
  \frac{\partial \rho}{\partial t} +
  \nabla\rho\vec{v}=0,
\end{equation}

\begin{equation}\label{mouv}
\frac{\partial \vec{v}}{\partial t} +
\vec{v}\cdot\nabla\cdot{v}=-\frac{\nabla
P_T}{\rho}+\frac{\vec{B}\cdot\nabla\vec{B}}{\mu_o\rho},
\end{equation}

\begin{equation}\label{ind}
  \frac{\partial \vec{B}}{\partial t}=
  \nabla\times(\vec{v}\times\vec{B}),
\end{equation}

\begin{equation}\label{pres}
  P= K \rho^\gamma,
\end{equation}

\noindent with standard notations, and where $P_T=P+B^2/2\mu_o$ is
the total (gas and magnetic) pressure, $K$ a constant, and
$\gamma$ the polytropic index.

\subsection{Equilibrium}

Using a cylindrical coordinate system ($r,\theta,z$), a static
($\mathbf{v}=0$) cylindrical column of axis $z$ is described by a
helical magnetic field $B_\theta(r), B_z(r)$, and a gas pressure
$P(r)$ depending only on the cylindrical radius $r$. The
continuity and induction equations are then trivially satisfied,
as well as the vertical and azimuthal component of the momentum
equation, while the radial component reduces to

\begin{equation}\label{radequi}
 - \frac{d P_T}{dr} -
\frac{B^2_{\theta}} {\mu_o r} = 0.
\end{equation}

This cylindrical equilibrium is best characterized by introducing
a number of quantities homogeneous to an inverse length, both in
vectorial ($\vec{\kb}$, $\vec{\kp}$ and $\vec{\kc}$) or algebraic
form ($\kb$, $\kp$ and $\kc$). They are defined by:

\begin{equation}\label{kb}
\vec{\kb}\equiv\frac{\vec{\nabla} B_o}{B_o}=
\frac{1}{B_o}\frac{dB_o}{dr}\mathbf{e}_r\equiv \kb\mathbf{e}_r,
\end{equation}

\begin{equation}\label{kp}
\vec{\kp}\equiv\frac{\vec{\nabla} P_o}{P_o}= \frac{1}{P_o}\frac{d
P_ o}{dr}\mathbf{e}_r\equiv \kp\mathbf{e}_r,
\end{equation}

\begin{equation}\label{kc}
\vec{\kc}\equiv\evec_\parallel \cdot\vec{\nabla}\evec_\parallel
=-\frac{B_\theta^2}{rB_o^2}\evec_r\equiv \kc \evec_r.
\end{equation}

\noindent where $B_o$ and $P_o$ are the equilibrium distribution
of magnetic field and gas pressure, and
$\vec{e}_\parallel=\vec{B}_o/B_o$ is the unit vector parallel to
the magnetic field; $\vec{\kc}$ is the curvature vector of the
magnetic field lines, and $\vec{\kb}$ characterizes the inverse of
the spatial scale of variation of the magnetic field, while
$\vec{\kp}$ characterizes the inverse scale of variation of the
fluid pressure. The first identity in these relations is general,
whereas the second one pertains to cylindrical equilibria only.

It is also convenient to introduce the plasma $\beta$ parameter:

\begin{equation}\label{beta}
\beta\equiv \frac{2\mu_o P_o}{B_o^2},
\end{equation}

\noindent This parameter measures the relative importance of the
gas and magnetic pressures.

With these definitions, the jet force equilibrium relation reads

\begin{equation}\label{kckb}
\frac{\beta}{2}\kp=\left( \kc -\kb \right),
\end{equation}

Both forms of the equilibrium relation, Eqs.~(\ref{radequi}) and
(\ref{kckb}), express the fact that the hoop stress due to the
magnetic tension ($\kc$) balances the gas ($\kp$) and magnetic
($\kb$) pressure gradient to achieve equilibrium and confine the
plasma in the column. Self-confinement is achieved in this way
when the external pressure is negligible at the column boundary.

\subsection{Perturbations:}

We want to investigate the stability with respect to deviations
from equilibrium. As the background equilibrium is static, the
problem is most easily formulated and analyzed in Lagrangian form:
indeed, in this case, all equations but the momentum equation can
be integrated with respect to time. To this effect, we introduce,
for any fluid particle at position $\vec{r}_o$ in the absence of
perturbation, the displacement $\vec{\xi}(\vec{r}_o,t)$ at time
$t$ from its unperturbed position, so that its actual position is
given by

\begin{equation}\label{displace}
  \vec{r}(\vec{r}_o,t)= \vec{r}_o+\vec{\xi}(\vec{r}_o,t).
\end{equation}

\noindent The unperturbed position $\vec{r}_o$ is used to uniquely
label all fluid elements.

Denoting by $\delta X$ the (Lagrangian) variation during the
displacement of any quantity $X$, the linearized (Eulerian)
equation of continuity $\partial_t \delta\rho=-\nabla(\rho_o
\vec{v})$ integrates into\footnote{In these expressions, the
difference between the Eulerian and Lagrangian variations has been
ignored as they disappear to first order in the displacement
$\vec{\xi}$ in the final equations. For the same reason, no
distinction is made between the derivative with respect to
$\vec{r}$ or $\vec{r}_o$.}

\begin{equation}\label{drho}
  \delta\rho=-\nabla(\rho_o\vec{\xi}).
\end{equation}

Similarly, the linearized induction equation $\partial_t
\delta\vec{B}=\nabla\times(\vec{B}_o \times \vec{v})$ leads to

\begin{equation}\label{dB}
  \delta\vec{B}=\nabla\times(\vec{B}_o \times \vec{\xi}).
\end{equation}

From these results and the polytropic equation of state, the total
pressure variation reads

\begin{equation}\label{dpress}
  \delta P_T=-\vec{\xi}\cdot\nabla P_o -\gamma P_o\nabla\cdot\vec{\xi} +
\frac{\vec{B}_o\cdot\delta\vec{B}}{\mu_o}.
\end{equation}

For static equilibria, one can without loss of generality take a
Fourier transform the linearized momentum equation with respect to
time. For a given Fourier mode, one can write
$\vec{\xi}(\vec{r},t)=\vec{\xi}(r)\exp i\omega t$, so that the
linearized momentum equation becomes

\begin{equation}\label{fourier}
-\rho_o\omega^2 \vec{\xi}=-\nabla \delta
P_T+\delta\vec{T}\equiv\vec{F}(\vec{\xi}),
\end{equation}

\noindent where
$\delta\vec{T}=(\vec{B}_o.\nabla\vec{B}+\vec{B}.\nabla\vec{B}_o)/\mu_o$
represents the variation of the magnetic tension
force\footnote{Within a factor $\rho_o$.}. The last identity in
Eq.~(\ref{fourier}) defines the linear operator $\vec{F}$,
operating on $\vec{\xi}$ through Eqs.~(\ref{dB}) and
(\ref{dpress}).

\section{The Energy Principle and its consequences:}\label{EnPrin}

The linear operator $\vec{F}$ of Eq.~(\ref{fourier}) is
self-adjoint, i.e., taking into account that $\vec F$ is real:

\begin{equation}\label{autoadj}
  \int{\vec\eta}\cdot
{\bf F}({\vec\xi})d^3 r = \int{\vec\xi}\cdot {\bf
F}({\vec\eta})d^3 r.
\end{equation}

\noindent A demonstration of this relation can be found, e.g., in
Freidberg \cite{F87} (cf p.~242 and Appendix A of the book).

As a consequence of this property of $\vec{F}$, an Energy
Principle can be formulated. Defining

\begin{equation}\label{dW}
\delta W({\vec\xi}^*,{\vec\xi})=-\frac{1}{2}\int{\vec\xi}^*\cdot
{\bf F}({\vec\xi})d^3 r,
\end{equation}

\noindent and

\begin{equation}\label{K}
 K({\vec\xi^*},{\vec\xi})=\frac{1}{2}\int\rho|{\vec\xi}|^2 d^3 r,
\end{equation}

\noindent and taking the scalar product of Eq.~(\ref{fourier})
with ${\vec\xi}^*$ leads to

\begin{equation}\label{omtot}
\omega^2=\frac{\delta W}{K}.
\end{equation}

\smallskip

The self-adjointness of ${\bf F}$ has two important consequences
(Energy Principle):

\begin{itemize}
  \item $\omega^2$ is also extremum with respect to a variation of
$\vec\xi$.
  \item Stability follows if and only if $\delta W > 0$ for all possible
$\vec\xi$.
\end{itemize}

\noindent Ascertaining stability through the last statement is
usually an impossible task. Instead, one usually makes use of the
Energy Principle in a less ambitious manner: if one can find some
displacement making $\delta W < 0$ then one has a sufficient
condition of instability (or, taking the converse statement, a
necessary condition of stability). This is actually how Suydam
criterion is demonstrated. First the expression of $\delta W$ is
simplified by taking advantage of the cylindrical geometry and
focusing on marginal stability and incompressible displacements
(as they make $\delta W$ more easily negative; see below). Next,
one chooses a particular form of displacement in the vicinity of
the magnetic resonance of an ($m,k$) mode, and looks under which
conditions this displacement makes $\delta W$ negative; the
condition turns out to be Suydam criterion for a well-chosen
displacement. These computations are rather lengthy and the reader
is referred to Freidberg's book \cite{F87} for details.

A useful form\footnote{The boundary term is ignored, as it is not
required in this discussion.} of $\delta W$ has been derived by
Bernstein \textit{et al.\ } \cite{BFKK58}, which reads (see
Freidberg \cite{F87}, p.~259)

\begin{eqnarray}\label{dWstand}
  \delta W = \frac{1}{2}\int & d^3\vec{r}\ \left[\frac{|\vec{Q}_\perp|^2}{\mu_o}+
  \frac{B_o^2}{\mu_o}|\nabla\cdot\vec{\xi}_\perp + 2
  \vec{\kc}\cdot\vec{\xi}_\perp|^2
  + \gamma P_o|\nabla\cdot\vec{\xi}|^2\right.
  \nonumber\\ & - \left.
  2P_o(\vec{\kp}\cdot\vec{\xi}_\perp)(\vec{\kc}\cdot\vec{\xi}^*_\perp)
  - J_\parallel(\vec{\xi}^*_\perp\times \vec{e}_\parallel)
  \cdot\vec{Q}_\perp\right],
\end{eqnarray}

\noindent where $\vec{\xi}_\perp$ is the component of the
displacement perpendicular to the unperturbed field $\vec{B}$,
$\vec{Q}_\perp = \nabla\times(\vec{\xi}_\perp\times\vec{B}_o)$ is
the perturbation in the magnetic field, $\vec{\kc}$ is the
curvature vector of the magnetic field and $\vec{\kp}$ the inverse
pressure length-scale vector defined earlier; $J_\parallel$ and
$\vec{e}_\parallel$ are the current and unit vector parallel to
the magnetic field, respectively. The quantities $\vec\kp$ and
$\vec\kc$ are defined in Eqs.~(\ref{kp}) and (\ref{kc}).

The first term describes the field line bending energy; it is the
term responsible for the propagation of Alfv\'en waves, through
the restoring effect of the magnetic tension, which makes field
lines acting somewhat like a rubber band. The second term is the
energy in the field compression, while the third is the energy in
the plasma compression. The fourth term arises from the
perpendicular current (as ${\bf\nabla}P={\bf J}_\perp\times{\bf
B}$ in a static equilibrium), and the last one arises from the
parallel current $J_\parallel$. Only these two terms can be
negative, and give rise to an instability if they are large enough
to make $\omega^2 < 0$. Pressure-driven instabilities are driven
by the first of these two terms, while current-driven
instabilities are due to the second one. Pressure-driven
instabilities are further subdivided into interchange and
ballooning modes, depending on the shape of the perturbation, but
the basic properties of these different modes are similar, and
this distinction will not be discussed further in these
notes\footnote{In particular, Suydam criterion applies also to
ballooning mode in cylindrical geometry; see Freidberg's book
\cite{F87}, pp.~401-402 for details}.

For our purposes here, we are mostly interested in what can be
learned from the form of the fourth term. First note that this
term is destabilizing in cylindrical geometry when $\kc \kp
> 0$; this justifies the necessary condition of instability
given in section \ref{sec:base}. Furthermore, Eq.~(\ref{omtot})
and (\ref{dWstand}) imply that the pressure-driving term produces
an inverse growth rate $\gamma$, of order of magnitude

\begin{equation}\label{gamma}
  |\gamma|^2 \sim C_S^2 \kc\kp\sim C_S^2/R_o^2,
\end{equation}

\noindent where $C_S$ is the sound speed and $R_o$ the column
radius. This is quite fast, comparable to the Kelvin-Helmholtz
growth rate in YSO jets. This order of magnitude will be used in
the next section to set up an ordering leading to analytically
tractable dispersion relations for pressure-driven unstable modes.

\section{Dispersion relation in the large azimuthal field
limit:}\label{disprel}

Most of the general results on pressure-driven instabilities were
obtained in the fusion literature either from the use of the
Energy Principle, or from the so-called Hain-L\"ust equation (a
reduced perturbation equation for the radial displacement
\cite{HL58} \cite{G71}). These approaches are quite powerful, but
not familiar to the astrophysics community, and involve a lot of
prerequisite.

It is more common in astrophysics to grasp the properties of an
instability through the derivation of a dispersion relation. There
are actually two papers doing this in the jet context for
pressure-driven instabilities; however, the first one, by Begelman
\cite{B98}, focuses on the relativistic regime which brings a lot
of added complexity to the discussion, and the second one
\cite{KLP00} is partially erroneous.

Fortunately, in the limit of a near toroidal field of interest
here, a dispersion relation can be derived ab initio by elementary
means, and this approach is adopted here. To this effect, it is
first useful to reexamine the behavior of the three MHD modes in
an homogeneous medium, in the limit of quasi-perpendicular
propagation. It is known that this limit allows the use of a kind
of WKB type of approach in the study of interchange and ballooning
pressure-driven modes (see, e.g., Dewar and Glasser \cite{DG83}),
a feature we shall take advantage of in these notes.

\subsection{MHD waves in quasi-perpendicular propagation in
homogeneous media:}

We consider an homogeneous medium pervaded by a constant magnetic
field ${\vec B}_o$. The analysis of linear perturbations in such a
setting leads to the well-known dispersion relation of the slow
and fast magnetosonic modes and the Alfv\'en mode. Our purpose
here is to point out useful features of these modes when the
wavevector is nearly perpendicular to the unperturbed magnetic
field.

\begin{figure}
  \centerline{\includegraphics{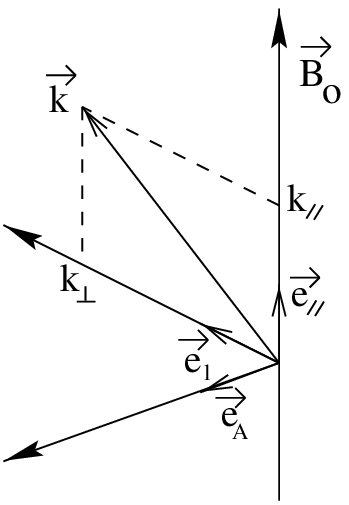}}
  \caption{Definition of the reference frame $(\ep,\el,\ea)$}
  \label{fig:triedre}
\end{figure}

To this effect, let us consider plane wave solutions to
Eq.~(\ref{fourier}), where $\xi\propto
\exp(-i\vec{k}\cdot\vec{r})$, and assume that the direction of
propagation is nearly perpendicular to the magnetic field, i.e.
$k_\parallel\ll k_\perp$ (defined in Eqs.~\ref{kpar} and
\ref{kperp}). The focus on quasi-perpendicular propagation comes
from the remarks of section \ref{suyd}, where it was noted that
instability is easier to achieve in the vicinity of magnetic
resonances, i.e., where $k_\parallel\ll k_\perp$.

Let us also introduce the orthogonal reference frame ($\ep$,
$\el$, $\ea$) where $\ep\equiv \vec{B}_o/B_o$ is parallel to the
unperturbed magnetic field, $\el\equiv\kpe/k_\perp$, and
$\ea\equiv\ep\times\el$ (see Fig.~\ref{fig:triedre}). With our
definition of $k_\parallel$ and $k_\perp$ in Eqs.~(\ref{kpar}) and
(\ref{kperp}), $\ea=\vec{e}_r$. The subscripts $l$ and $A$ stand
for longitudinal and alfv\'enic, respectively ($\ep$, $\el$, and
$\ea$ are the directions of the displacement of purely slow, fast
and alfv\'enic modes in the limit of nearly transverse propagation
adopted here, as shown below).

Denoting $(\xi_\parallel,\xi_l,\xi_A)$ the components of the
lagrangian displacement $\vec\xi$ in this reference frame, the
momentum equation Eq.~(\ref{fourier}) yields the following three
component equations

\begin{equation}
  \left( \omega^2 - C_S^2\, k_\parallel^2 \right) \, \xi_\parallel = C_S^2 \,
k_\parallel \, k_\perp \, \xi_l,
  \label{equ:mgs1}
\end{equation}

\begin{equation}
  \left( \omega^2 - C_S^2\, k_\perp^2 - V_A^2\, k^2 \right) \,\xi_l = C_S^2
  \, k_\parallel \,k_\perp \,\xi_\parallel,
  \label{equ:mgs2}
\end{equation}

\begin{equation}
  \left( \omega^2 - V_A^2\,k_\parallel^2 \right) \,\xi_A = 0,
  \label{equ:alf}
\end{equation}

\noindent while the total pressure perturbation becomes

\begin{equation}
 \label{equ:dp}
    \delta P_T  =-i\rho_o \left[ (C_S^2+V_A^2) \,k_\perp\xi_l +
    C_S^2\,k_\parallel\xi_\parallel \right].
\end{equation}

Eq.~(\ref{equ:alf}) gives the dispersion relation of Alfv\'en
waves, $\omega_A^2=V_A^2\kpa^2$, which decouple from the two
magnetosonic modes described by the remaining two equations. The
solutions of the magnetosonic modes are easily derived and possess
the following important properties. Characterizing
quasi-perpendicular propagation with the small parameter
$\epsilon\equiv |k_\parallel/k_\perp|\ll 1$, these two equations
imply $\omega_S^2\simeq C_S^2 V_A^2/(C_S^2+V_A^2) k_\parallel^2$
and $\xi_l\sim O(\epsilon\xi_\parallel)$ for the slow magnetosonic
wave, while $\omega_F^2\simeq(C_S^2+V_A^2) k_\perp^2$ and
$\xi_\parallel\sim O(\epsilon\xi_l)$ for the fast magnetosonic
one.

Furthermore, the $\xi_l$ momentum component Eq.~(\ref{equ:mgs2})
combined with Eq.~(\ref{equ:alf}) and the ordering of the
displacement component just pointed out implies that $\delta
P_T=0$ to leading order in $\epsilon$ for the slow magnetosonic
mode; note that the same property holds by construction for the
Alfv\'en mode. The cancellation of the total pressure for these
two modes is essential from a technical point of view, and will
lead to substantial simplification in the derivation of a
dispersion relation performed in the next subsection.

\subsection{Dispersion relation and Kadomtsev criteria:}

Let us now come back to cylindrical inhomogeneous equilibria.
Remember from section \ref{EnPrin} that the pressure-driving term
will contribute a destabilizing term $\omega^2\sim C_S^2/R_o^2$ to
the dispersion relation. This term will be able overcome the
stabilizing effect of the restoring forces of the Alfv\'en and
slow magnetosonic modes only if $V_A |k_\parallel|,\ C_S
|k_\parallel| \lesssim C_S/R_o$. This constraint can be achieved
in the vicinity of magnetic resonance as already previously noted.

More precisely, a simplified dispersion relation can be found in
the WKB limit with a displacement of the form
$\vec{\xi}(\vec{r})=\vec{\xi}\times\exp - i(k_r r + m\theta+ k_z
z)$, if the following ordering is satisfied:

\begin{itemize}
  \item $|k_\parallel r|\ll$ or $\lesssim 1\ll |k_r r|\ll |k_\perp r|$:
the first inequality ensures that the stabilization by magnetic
tension is ineffective (closeness to a resonance). The following
inequalities ensure that a WKB limit can be taken. The implied
ordering\footnote{For consistency with the previous sections,
$k_\perp$ is the wavenumber in the longitudinal direction; it does
not include the piece in the radial direction.} $|k_\parallel|\ll
k_\perp$ ensures that $\delta P^*$ will vanish to leading order as
in the homogeneous case discussed in the previous section. The
last inequality allows us to neglect the contribution of the
radial gradient of total pressure (which \textit{does not}
vanish), and greatly simplifies the analysis.
  \item $|B_z/B_\theta|^2 s^2 |k_\perp|\ll |k_\parallel|$: this
limit, which applies when $|B _\theta| \gg |B_z|$, ensures that
the magnetic shear is not stabilizing.
  \item $|\omega^2|\ll |\omega_F|^2$: this excludes the fast
mode from the problem in the near perpendicular propagation regime
considered here. As the fast mode is not expected to be
destabilized in this regime (as $|\omega_F^2|\gg V_A^2/r^2$), this
does not limit the generality of the results while simplifying the
analysis.
\end{itemize}

It turns out that the resulting dispersion relation captures most
of the physics of pressure-driven instabilities; this follows
because the most unstable modes have growth rates nearly
independent of the azimuthal wavenumber $m$ \cite{DTYNM04}, and
because current-driven instabilities are efficient only at low $m$
and disappear from a WKB analysis.

As previously, the projection Eq.~(\ref{fourier}) on the
longitudinal direction $\el$ shows that the total pressure
perturbation vanishes and that $\xi_l\sim |k_\parallel/k_\perp|
\xi_\parallel \ll \xi_\parallel$), while the components in the
other two directions ($\ep, \er$) are now coupled and read (some
details of the derivation of these equations can be found in
Appendix \ref{details})

\begin{equation}\label{slow}
  \left(\omega^2 - V_{SM}^2 k_\parallel^2\right)=-i\frac{2\beta^*}{1+\beta^*}
V_A^2\kc k_\parallel \xi_r,
\end{equation}

\begin{equation}\label{alfven}
  \left[\omega^2 - V_A^2 (k_\parallel^2+k_o^2)\right]=i\frac{2\beta^*}{1+\beta^*}
V_A^2\kc k_\parallel \xi_\parallel,
\end{equation}

\noindent where $\beta^*=C_S^2/V_A^2$, and $V_{SM}^2=C_S^2
V_A^2/(C_S^2+V_A^2)$ is the slow mode speed in the near
perpendicular propagation limit. The coupling of the modes blurs
their character except in limiting cases.

The quantity $k_o^2$ is defined as

\begin{equation}\label{kzero}
  k_o^2=\frac{4\beta^*}{1+\beta^*}\kc^2 - 2\beta^* \kc\krho.
\end{equation}

Note that if $\kc=0$ (i.e., when reverting to an homogeneous
medium), Eqs.~(\ref{slow}) and (\ref{alfven}) yield back the slow
and Alfv\'en mode, respectively. The field curvature couples the
two modes. The quantity $k_o^2$ can be either positive or
negative; the first term in Eq.~(\ref{kzero}) comes from the
plasma compression, and the second one is the contribution of the
pressure destabilizing term identified in section \ref{EnPrin}.

As usual, these equations possess a non-trivial solution if their
determinant is non zero, which yields the following dispersion for
$\omega^2$:

\begin{equation}\label{disp}
  \omega^4 - \left[(V_A^2+V_{SM}^2) k_\parallel^2 + V_A^2
k_o^2\right]\omega^2 + V_A^2V_{SM}^2 k_\parallel^2 (k_\parallel^2
-2\beta^*\kc\krho)=0.
\end{equation}

First note that if both $B_z=0$ (the so-called Z-pinch
configurations) and $m=0$, this equation is degenerate: one of the
roots is $\omega^2=0$ and the other root is $\omega^2=V_A^2
k_o^2$. Instability then requires that $k_o^2<0$, as
$k_\parallel=0$ in this case. This constrain is identical to the
criterion\footnote{Kadomtsev's criterion for the $m=0$ mode in a Z
pinch is a necessary and sufficient condition of instability,
whereas the analysis presented here shows only the sufficiency of
this condition.} derived by Kadomtsev from the Energy Principle
for the $m=0$ mode in Z pinches (see Freidberg \cite{F87} p.\
286).

When $m \neq 0$, Eq.~(\ref{disp}) can be solved exactly but it is
more instructive to analyze its properties. As the coefficient of
$\omega^2$ is equal to the sum of the two roots, and the last term
is equal to their product, one finds that if $k_\parallel^2> 2
\beta^*\kc\krho$, the two roots are stable, and if $k_\parallel^2
< 2 \beta^*\kc\krho$, one of the roots is unstable. If $B_z=0$ (Z
pinch), this condition is identical to the criterion\footnote{Same
comment as in the previous footnote.} derived by Kadomtsev for
$m\neq 0$ modes (see Freidberg \cite{F87} pp.\ 284-285).

Note that all these conditions of instability require $\kc\kp
> 0$, in agreement with the discussion of sections \ref{EnPrin} and
\ref{sec:gen}; this condition is unavoidable in magnetically
self-confined jets. The analysis presented here also shows that
once this condition is satisfied, instability necessarily follows
in static columns where $|B_\theta|\gg |B_z|$ on some of the
radial range, as the magnetic tension stabilizing effect $V_A^2
k_\parallel^2$ is arbitrarily small in the vicinity of a magnetic
resonance.

Finally, the reader may ask how the local analysis presented here
informs us on the global stability properties of the column. The
answer lies in in oscillation theorem of Goedbloed and Sakanaka
\cite{GS74}. The theorem states that for any ($m,k$) unstable
mode, the growth rate decreases when increasing the number of
radial nodes. This implies that if an unstable mode with a large
number of radial nodes is found (such as the modes considered
here), an unstable nodeless mode will also exist, and this mode
will have the largest growth rate. Such a mode will have a very
disruptive effect on the plasma if its displacement is not
vanishing on the boundary, as will be the case if the azimuthal
field is dominant on the boundary.

\section{Moving columns:}\label{movcol}

The previous section has shown that cylindrical columns with a
predominant azimuthal magnetic field at least in some radial range
are subject to pressure-driven instabilities. This situation holds
in the outer region of self-confined magnetic jets, leading to a
potentially disruptive configuration. However, in these regions a
gradient of axial velocity due to the interaction of the moving
jet with the outside medium is also expected to be present, and it
is legitimate to investigate the effect of such a velocity
gradient on the stability properties of pressure-driven modes.

This problem has not yet been addressed in the astrophysics
literature, but some relevant results are available in the fusion
literature. In all the investigations cited below, the adopted
velocity profile contains no inflexion point, in order to avoid
the triggering of the Kelvin-Helmhotz instability.

It is first useful to consider what becomes of Suydam criterion in
presence of background motions\footnote{This requires a
generalization of Eq.~(\ref{fourier}); also, the Energy Principle
no longer applies as the resulting operator is not self-adjoint.}.
This investigation has been performed by Bondeson \etal
\cite{BIB87}. Focusing on axial flows (${\bf U}=U_z(r) \er$), they
conclude that the behavior of localized modes depends on the
magnitude of

\begin{equation}\label{shear-Mach}
  M\equiv \rho^{1/2}\frac{U'_z}{q' B_z/q},
\end{equation}

\noindent where the prime denotes radial derivative, and $q$ is
the safety factor (see section \ref{idealMHD}). This quantity is a
form of Alfv\'enic Mach number based on the velocity and magnetic
shear, hence its name. When $M^2 < \beta$, the flow shear
destabilizes resonant modes. Above this limit, these modes are
stable, but in this case, unstable modes exist at the edge of the
slow continuum, and may be global. The authors found however that
in this case the growth rates are small (comparable to the
resistive instabilities growth rates). Note also that, as
$q'/q\sim 1/r$, $M\sim (B_\phi/B_z)(r/d)(U_z/V_A) \gg 1$ in MHD
jets ($d$ is the width of the velocity layer).

These results seem to suggest that the region where the velocity
shear layer takes place at the jet boundary is substantially
stabilized in MHD jets. This seems to be confirmed by global
linear stability analyzes, both for interchange and ballooning
modes, except possibly for the $m=0$ (``saussage") mode \cite{C96}
\cite{SH95} \cite{WC91}. In all cases, increasing the flow Mach
number efficiently reduces the amplitude of the displacement of
the unstable modes at the plasma boundary, an important feature to
avoid the disruption of the plasma.

An efficient stabilization mechanism has also been identified in
the nonlinear regime by Hassam \cite{H92}. This author exploits an
analogy between the $m=0$ pressure-driven interchange mode and the
Rayleigh-Taylor instability in an appropriately chosen magnetized
plasma configuration. From this analysis, he concludes that the
$m=0$ pressure-driven mode is nonlinearly stabilized by a smooth
velocity shear ($dU_z/dr\sim U/R_o$) if $M_s=U_z/C_S \gtrsim [\ln
(\tau_d/\tau_g)]^{1/2}$, where $\tau_g$ is the instability growth
time-scale ($\tau_g\sim c_s(\krho\kc)^{1/2}$) and $\tau_d$ the
diffusion time-scale ($\tau_d\sim\nu\krho\kc$ where $\nu$ is the
viscosity, assumed comparable to the resistivity). The nonlinear
evolution of an unstable, slightly viscous and resistive Z-pinch
(i.e., a configuration where the field is purely azimuthal), was
simulated by Desouza-Machado \etal \cite{DHS00}. They found that
the plasma relaminarizes over almost all its volume for applied an
velocity shear in good agreement with this analytic estimate. The
core of the plasma still has some residual unstable ``wobble",
which can apparently be stabilized by the magnetic shear if some
longitudinal field $B_z$ is added to the configuration. Note that
the large values of $\tau_d/\tau_g$ relevant to astrophysical jets
lead to only weak constraints\footnote{For example
$\tau_d/\tau_g=10^{30}$ translates into $M_s \gtrsim 8$ only; in
YSO jets, this ratio is most probably significantly smaller, and
the constraint even weaker.} on the Mach number $M_s$, so that
this nonlinear stabilization mechanism is expected to be efficient
in astrophysical jets.

\section{Summary and open issues:}

Pressure-driven instabilities occur in static columns when the
pressure force pushes the plasma out from the inside of the
magnetic field lines curvature, as shown from direct inspection of
the ``potential energy" of the linearized displacement equation
(section \ref{EnPrin}), and from a the dispersion relation of
local modes (section \ref{disprel}). When unstable modes exist,
the growth rates are of the order $C_S/R_o$ where $C_S$ is the
sound speed and $R_o$ the jet radius. These are very large,
comparable to the Kelvin-Helmholtz growth rate (the most studied
instability in jets), especially that the ratio of the magnetic
energy to the gas internal energy is expected to be of order unity
(within an order of magnitude or so). Such instabilities are known
to be disruptive in the fusion context when the eigenmodes exhibit
a substantial displacement of the plasma outer boundary; such a
situation is relevant to magnetically self-confined jets, as the
magnetic field in their outer region is predominantly azimuthal, a
configuration most favorable to the onset of the instability
(sections \ref{sec:gen} and \ref{disprel}). However, the presence
of a velocity gradient in the outer boundary due to the jet bulk
motion is expected to have a substantial stabilizing influence,
both in the linear and nonlinear regimes (section \ref{movcol}).

In its present state, this picture possesses a number of loose
ends:

\begin{itemize}
  \item The stabilizing role of an axial velocity gradient needs
to be better understood. Not all modes may be stabilized in the
linear regime, depending on the details of the equilibrium jet
configuration, and the nonlinear mechanism identified in the
literature is highly idealized and may not be generic. The one and
only simulation of nonlinear stabilization published to date
exhibits a very violent relaxation transient, which may still lead
to jet disruption. On the other hand, this transient is also an
indication that the initial configuration of the simulation is way
out of equilibrium, a situation which may not occur in real jets.
  \item The role of jet rotation has not yet been correctly
investigated. Preliminary results seem to indicate that it is
stabilizing \cite{KLP00}; however, jet rotation may not be an
important dynamical factor in the asymptotic jet propagation
regime.
  \item Most investigations of pressure-driven instabilities rely
on a very simple prescription of the equation of state, which
raises an issue of principle. Indeed, the very large growth rates
usually found for the instability indicate that it develops on
time-scales much shorter than the collisional time-scale, and the
use of ideal MHD as well as a polytropic equation of state may be
questioned in such a context, an issue briefly addressed in
Appendix \ref{app:idealmhd}. A more complex description of the
plasma is required to validate the results obtained so far.
\end{itemize}

\appendix

\section*{Appendices}

\renewcommand{\thesection}{\Alph{section}}
\setcounter{section}{0}

\section{On the use of ideal MHD:} \label{app:idealmhd}

In astrophysics in general, and jet stability analyzes in
particular, an MHD framework is almost always adopted instead of
the more precise kinetic one, due to its relative simplicity. The
MHD approximation can be applied when the fluid is locally
neutral, when all species can be described by a single fluid
equation (i.e. when the relative drift velocity of species with
respect to one another is small), and when Ohm's law is valid. The
validity of these approximations has been discussed elsewhere
\cite{KT86} \cite{S92} and will not be reproduced here; the
interested reader is referred to these books for details.

Furthermore, MHD stability (and jet stability in particular) is
often investigated within the framework of ideal MHD. Indeed, the
dynamical time-scales of interest (including those of the
considered instability) is almost always substantially larger than
the particle collision time-scale. Moreover, an isotropic pressure
is often assumed, e.g. through a barotropic equation of state, and
this raises another issue of principle within the framework of
ideal MHD, as, indeed, an isotropic pressure would be expected
only if collisions at the particle level are not neglected.

The isotropic pressure assumption can be justified to some extent
by the fact that plasma microturbulence does limit pressure
anisotropy to a factor of order unity. For example, within the
framework of collisionless MHD, pressure anisotropy is
self-limiting (for a recent synthetic discussion of this problem
within the framework of collisionless magnetorotational
instability, see Sharma \etal \cite{SHQS06} and references
therein). Nevertheless, this provides little support (if any) to
the adoption of a closure in the form of a barotropic or adiabatic
equation of state in a collisionless setting.

Collisionless MHD approximations apply when the length-scales and
frequencies under consideration are larger than the ion Larmor
radius, and smaller than the ion cyclotron frequency,
respectively. These conditions should be satisfied in jets, but I
am not aware of any investigation of pressure-driven instabilities
in this framework. Freidberg \cite{F87} argues that a simple rule
of thumb to estimate the effects of the assumed closure is to
replace the adiabatic index by $0$, and to assume
incompressibility of the motions within the framework of standard
ideal MHD.

\section{Derivation of the dispersion relation:}\label{details}

Some intermediate steps in the derivation of the dispersion
relation of section \ref{disprel} are given here. The notations
are the same as in this section.

Direct computation of the pressure and magnetic field perturbation
gives

\begin{equation}\label{dp}
  \delta P = -\rho_o C_S^2\left[\vec{\nabla}\cdot \vec{\xi} +
\krho \xi_r\right],
\end{equation}

\begin{eqnarray}\label{dbfield}
  \delta \vec{B} = - \vec{B}_o &
\left[\vec{\nabla}\cdot(\vec{\xi}_r+\vec{\xi}_l)+\xi_r(\kb+\kc)\right]\ep
\nonumber\\ & -i k_\parallel B_o (\vec{\xi}_r+\vec{\xi}_l).
\end{eqnarray}

This allows us to obtain the perturbation in total pressure and in
magnetic tension:

\begin{eqnarray}\label{dptot}
  \delta P_T = -\rho_o & (V_A^2 + C_S^2 ) \vec{\nabla}\cdot
(\vec{\xi}_r + \vec{\xi}_l) + i \rho_o C_S^2 k_\parallel
\xi_\parallel - \nonumber\\ & \rho_o
\vec{\xi}_r\cdot(V_A^2\vec{\kb}+V_A^2\vec{\kc}+C_S^2\vec{\krho}),
\end{eqnarray}

\begin{eqnarray}\label{dT}
  \delta \vec{T}= & -V_A^2 \vec{\kc}\left[2\vec{\nabla}\cdot(\vec{\xi}_r+\vec{\xi}_l)
+2 \vec{\xi}_r\cdot(\vec{\kb}+\vec{\kc})\right]+ \nonumber\\ & i
k_\parallel V_A^2\left[\vec{\nabla}\cdot (\vec{\xi}_r +
\vec{\xi}_l)+ 2\vec{\xi}_r\cdot\vec{\kc}\right]\ep - k_\parallel^2
V_A^2 (\vec{\xi}_r + \vec{\xi}_l).
\end{eqnarray}

Furthermore, the equilibrium relation Eq.~(\ref{kckb}) allows us
to eliminate $\kb$ in terms of $\kc$ and $\krho=\gamma\kp$.

The longitudinal component of the linearized momentum equation
reduces to $\delta P_T=0$, once contributions of order
$k_\parallel \xi_l$ or $\xi_l/r$ are neglected in front of
$k_\perp \xi_l$. This constraint shows that $\xi_l \sim
O(k_\parallel/k_\perp \xi_r, k_\parallel/k_\perp \xi_\parallel)$.
It also allows us to eliminate
$\vec{\nabla}\cdot(\vec{\xi}_r+\vec{\xi}_l)$ from the remaining
two component equations, which then reduce to Eqs.~(\ref{slow})
and (\ref{alfven}). In the process, the contribution of $d\delta
P_T/dr$ is shown to be negligible from the assumed ordering
relations $|k_r|\ll k_\perp$ and $|B_z/B_\theta|^2 s^2
|k_\perp|\ll |k_\parallel|$, i.e., the magnetic shear stabilizing
term can be neglected in this limit.

\renewcommand{\thesection}{\arabic{section}}
%
% BibTeX users please use
%
\bibliographystyle{spmpsci}
\bibliography{jet-biblio}

\begin{thebibliography}{10}
\providecommand{\url}[1]{{#1}}
\providecommand{\urlprefix}{URL }
\expandafter\ifx\csname urlstyle\endcsname\relax
  \providecommand{\doi}[1]{DOI~\discretionary{}{}{}#1}\else
  \providecommand{\doi}{DOI~\discretionary{}{}{}\begingroup
  \urlstyle{rm}\Url}\fi

\bibitem{ALB00}
{Appl}, S., {Lery}, T., {Baty}, H.: {Current-driven instabilities in
  astrophysical jets. Linear analysis}.
\newblock \aap \textbf{355}, 818--828 (2000)

\bibitem{BK02}
{Baty}, H., {Keppens}, R.: {Interplay between Kelvin-Helmholtz and
  Current-driven Instabilities in Jets}.
\newblock \apj \textbf{580}, 800--814 (2002)

\bibitem{B98}
{Begelman}, M.C.: {Instability of Toroidal Magnetic Field in Jets and
  Plerions}.
\newblock \apj \textbf{493}, 291 (1998)

\bibitem{BFKK58}
{Bernstein}, I.B., Frieman, E.A., Kruskal, M.D., Kulsrud, R.M.: \textit{Proc.\
  Roy.\ Soc.\ London A} \textbf{244}, 17 (1958)

\bibitem{Bir91}
Birkinshaw, M.: The stability of jets.
\newblock In: P.~Hughes (ed.) Beams and Jets in Astrophysics, pp. 278--341.
  Cambridge University Press (1991)

\bibitem{BR74}
{Blandford}, R.D., {Rees}, M.J.: {A 'twin-exhaust' model for double radio
  sources}.
\newblock \textit{Monthly Not.\ Royal Astron.\ Soc.} \textbf{169}, 395 (1974)

\bibitem{Bod95}
{Bodo}, G., {Massaglia}, S., {Rossi}, P., {Rosner}, R., {Malagoli}, A.,
  {Ferrari}, A.: {The long-term evolution and mixing properties of high Mach
  number hydrodynamic jets.}
\newblock \textit{Astron.\ Astrophys.} \textbf{303}, 281 (1995)

\bibitem{Bod98}
{Bodo}, G., {Rossi}, P., {Massaglia}, S., {Ferrari}, A., {Malagoli}, A.,
  {Rosner}, R.: {Three-dimensional simulations of jets}.
\newblock \textit{Astron.\ Astrophys.} \textbf{333}, 1117 (1998)

\bibitem{BIB87}
{Bondeson}, A., {Iacono}, R., {Bhattacharjee}, A.: {Local magnetohydrodynamic
  instabilities of cylindrical plasma with sheared equilibrium flows}.
\newblock Physics of Fluids \textbf{30} (1987)

\bibitem{C96}
{Chiueh}, T.: {Suppression of the edge interchange instability in a straight
  tokamak}.
\newblock Phys.\ Rev.\ E \textbf{54}, 5632--5635 (1996)

\bibitem{DHS00}
{Desouza-Machado}, S., {Hassam}, A.B., {Sina}, R.: {Stabilization of Z pinch by
  velocity shear}.
\newblock Physics of Plasmas \textbf{7}, 4632--4643 (2000)

\bibitem{DG83}
{Dewar}, R.L., {Glasser}, A.H.: {Ballooning mode spectrum in general toroidal
  systems}.
\newblock Physics of Fluids \textbf{26}, 3038--3052 (1983)

\bibitem{DTYNM04}
{Dewar}, R.L., {Tatsuno}, T., {Yoshida}, Z., {N{\"u}hrenberg}, C., {McMillan},
  B.F.: {Statistical characterization of the interchange-instability spectrum
  of a separable ideal-magnetohydrodynamic model system}.
\newblock Phys.\ Rev.\ E \textbf{70}, 066,409 (2004)

\bibitem{F98}
{Ferrari}, A.: {Modeling Extragalactic Jets}.
\newblock \araa pp. 539--598 (1998)

\bibitem{F97}
{Ferreira}, J.: {Magnetically-driven jets from Keplerian accretion discs.}
\newblock \aa \textbf{319}, 340--359 (1997)

\bibitem{F87}
{Freidberg}, J.P.: {Ideal Magnetohydrodynamics}.
\newblock Plenum Press (1987)

\bibitem{G71}
{Goedbloed}, J.P.: {Stabilization of magnetohydrodynamic instabilities by
  force-free magnetic fields. II. Linear pinch.}
\newblock Physica \textbf{53}, 501--534 (1971)

\bibitem{GS74}
{Goedbloed}, J.P., Sakanaka, P.H.: Phys.\ Fluids \textbf{17}, 908 (1974)

\bibitem{HL58}
{Hain}, K., {L\"ust}, R.: Z. Naturforsch. Teil A \textbf{13}, 936 (1958)

\bibitem{H92}
{Hassam}, A.B.: {Nonlinear stabilization of the Rayleigh-Taylor instability by
  external velocity shear}.
\newblock Physics of Fluids B \textbf{4}, 485--487 (1992)

\bibitem{KLP00}
{Kersal{\'e}}, E., {Longaretti}, P.Y., {Pelletier}, G.: {Pressure- and magnetic
  shear-driven instabilities in rotating MHD jets}.
\newblock \aa \textbf{363}, 1166--1176 (2000)

\bibitem{KT86}
{Krall}, N.A., {Trievelpiece}, A.W.: {Principles of Plasma Physics}.
\newblock San Francisco Press (1986)

\bibitem{L96}
{Li}, Z.: {Magnetohydrodynamic Disk-Wind Connection: Magnetocentrifugal Winds
  from Ambipolar Diffusion-dominated Accretion Disks}.
\newblock \apj \textbf{465}, 855--+ (1996)

\bibitem{RJF00}
{Ryu}, D., {Jones}, T.W., {Frank}, A.: {The Magnetohydrodynamic
  Kelvin-Helmholtz Instability: A Three-dimensional Study of Nonlinear
  Evolution}.
\newblock \apj \textbf{545}, 475--493 (2000)

\bibitem{SHQS06}
{Sharma}, P., {Hammett}, G.W., {Quataert}, E., {Stone}, J.M.: {Shearing Box
  Simulations of the MRI in a Collisionless Plasma}.
\newblock \apj \textbf{637}, 952 (2006)

\bibitem{S92}
{Shu}, F.H.: {Physics of Astrophysics, Vol. II}.
\newblock University Science Books (1992)

\bibitem{SH95}
{Shumlak}, U., {Hartman}, C.W.: {Sheared Flow Stabilization of the m = 1 Kink
  Mode in Z Pinches}.
\newblock Physical Review Letters \textbf{75}, 3285--3288 (1995)

\bibitem{WC91}
{Waelbroeck}, F.L., {Chen}, L.: {Ballooning instabilities in tokamaks with
  sheared toroidal flows}.
\newblock Physics of Fluids B \textbf{3}, 601--610 (1991)

\end{thebibliography}
%
%%%%%%%%%%%%%%%%%%%%%%%%%%%%%%%%%%%%%%%%%%%%%%%%%%%%%%%%%%%%%%%%%%%%%%
%
%%%%%%%%%%%%%%%%%%%%%%%%%%%%%%%%%%%%%%%%%%%%%%%%%%%%%%%%%%%%%%%%%%%%%%
%
%\printindex
\end{document}